\documentclass[12pt]{article}
\usepackage{epsf}
\title{Nucleon spin content and axial coupling
constants in QCD sum rules approach.}
\author{{\it Lecture at St.Petersburg Winter School
on Theoretical Physics,} \\
{\it  Febr. 23-28, 1998}\\
\\
B.L.Ioffe\\
Institute of Theoretical and Experimental Physics\\
117218, Moscow, Russia}
\date{}
\begin{document}
\maketitle

\newcommand{\be}{\begin{equation}}
\newcommand{\ee}{\end{equation}}

\def\la{\mathrel{\mathpalette\fun <}}
\def\ga{\mathrel{\mathpalette\fun >}}
\def\fun#1#2{\lower3.6pt\vbox{\baselineskip0pt\lineskip.9pt
\ialign{$\mathsurround=0pt#1\hfil##\hfil$\crcr#2\crcr\sim\crcr}}}


\vspace{1cm}

\begin{abstract}
The review of current experimental situation in the measurements of the
first moment $\Gamma_{p,n}$ of spin dependent nucleon structure functions
$g_{1;p,n}(x,Q^2)$ is presented. The results of the calculations of twist-4
corrections to $\Gamma_{p,n}$ are discussed and their accuracy is estimated.
The part of the proton spin $\Sigma$ carried by $u, d, s$ quarks is
calculated in the framework of the QCD sum rules in the external
fields. The operators up to dimension 9 are accounted.  An important
contribution comes from the operator of dimension 3, which in the limit
of massless $u, d, s$ quarks is equal to the derivative
of QCD topological  susceptibility $\chi^{\prime} (0)$. The comparison
with the experimental data on $\Sigma$ gives $\chi^{\prime}(0)= (2.3
\pm 0.6) \times 10^{-3} ~ GeV^2$. The limits on $\Sigma$ and
$\chi^{\prime}(0)$ are found from selfconsistency of the sum rule,
$\Sigma \ga 0.05,~~ \chi^{\prime} (0) \ga 1.6 \times 10^{-3} ~ GeV^2$. The
values of $g_A = 1.37 \pm 0.10$ and $g^8_A = 0.65 \pm 0.15$ are also
determined from the corresponding sum rules.
\end{abstract}

\newpage
{\it I dedicate this lecture to the memory of my friend Volodya Gribov, whom I
knew for about half a centure. Now it becomes even more clear how great was
his influence on physics: his brilliant ideas, his uncompromising approach
to science, his teaching ability. My loss is even more painful: every
meeting with Volodya was like a holyday to my soul.}


\vspace{7mm}


{\bf 1.~ Introduction. Recent experimental data.}

\vspace{2mm}

In the last years, the problem of nucleon spin content and particularly
the question which part of the nucleon spin is carried by quarks,
attracts a strong interest. The valuable information comes from the
measurements of the spin-dependent nucleon structure functions $g_1(x,
Q^2)$ in deep inelastic $e(\mu)N$ scattering (for the recent data see
[1,2,3], for a reviews [4,5]). The parts of the nucleon spin carried by
$u, d$
and $s$-quarks are determined from the measurements of the first moment
of $g_1(x, Q^2)$

\begin{equation}
\Gamma_{p,n}(Q^2) = \int \limits_{0}^{1} dx g_{1;p,n} (x, Q^2)
\end{equation}
At high $Q^2$ with the account of twist-4 contributions $\Gamma_{p,n}(Q^2)$
have the form

\be
\Gamma_{p,n}(Q^2) = \Gamma^{as}_{p,n}(Q^2) + \Gamma^{tw 4}_{p,n}(Q^2)
\ee
$$
\Gamma^{as}_{p,n}(Q^2) = \frac{1}{12} \Biggl \{ [1 - \bar{\alpha} - 3.58
\bar{\alpha}^2 - 20.2 \bar{\alpha}^3 - c \bar{\alpha}^4 ] [\pm g_A +
\frac{1}{3}g^8_A] $$

\be
+ \frac{4}{3} [1 - \frac{1}{3} \bar{\alpha} - 0.55 \bar{\alpha}^2 - 4.45
\bar{\alpha}^3] \Sigma \Biggr \} -
\frac{N_f}{18 \pi} \alpha_s(Q^2) \Delta g(Q^2)
\ee

\be
\Gamma^{tw4}_{p,n}(Q^2) = \frac{b_{p,n}}{Q^2}
\ee
In eq.(3) $\bar{\alpha} = \alpha_s(Q^2)/\pi, g_A$ is the $\beta$-decay axial
coupling constant, $g_A = 1.260 \pm 0.002$ [6]

\be
g_A = \Delta u - \Delta d ~~~~ g^8_A = \Delta u + \Delta d - 2 \Delta s
~~~~ \Sigma = \Delta u + \Delta d + \Delta s.  \ee $\Delta u, \Delta d,
\Delta s, \Delta g$ are parts of the nucleon spin projections carried by $u,
d, s$ quarks and gluons:

\be
\Delta q = \int \limits^{1}_{0} \Biggl [q_+(x) - q_-(x) \Biggr ]
dx
\ee
where $q_+(x), q_-(x)$ are quark distributions with spin projection
parallel (antiparallel) to nucleon spin and a similar definition takes place
for $\Delta g$. The coefficients of perturbative series were calculated in
[7-10], the numerical values in (3) correspond to the number of flavours
$N_f = 3$, the coefficient $c$ was estimated in [11], $c \approx 130$. In
the $\overline{MS}$ renormalization scheme chosen in [7-10]
$g_A,g^8_A$ and $\Sigma$ are
$Q^2$-independent. In the assumption of the exact $SU(3)$ flavour symmetry
of the octet axial current matrix elements over baryon octet states $g^8_A =
3F - D = 0.59 \pm 0.02$ [12].
On the basis of operator product expansion (OPE) the quantities $g_A, g^8_A$
and $\Sigma$ are related to the proton matrix element of isovector, octet
and singlet axial currents correspondingly:

\be
2ms_{\mu}(g_A,g_A^8,\Sigma) = \langle p,s \mid j^{(3)}_{\mu 5},
j^{(8)}_{\mu 5}, j^{(0)}_{\mu 5}\mid p,s \rangle,
\ee
where $s_{\mu}$ is the proton spin 4-vector, $m$  is the proton mass.

Strictly speaking, in (3) the separation of terms proportional to $\Sigma$
and $\Delta g$ is arbitrary, since OPE has
only one singlet in flavour twist-2 operator for the first moment of the
polarized structure function -- the operator of singlet axial current
$j^{(0)}_{\mu 5}(x) = \sum \limits_{q}~ \bar{q}_i(x) \gamma_{\mu} \gamma_5 q, ~~ q =
u,d,s$. The separation of terms proportional to $\Sigma$ and $\Delta g$ is
outside the framework of OPE and depends on the infrared cut-off. The
expression used in (3) is based on the physical assumption that the
virtualities $p^2$ of gluons in the nucleon are much larger than light quark
mass squares, $\vert p^2 \vert \gg m^2_q$ [13] and that the infrared cut-off is chosen
in a way providing the standard form of axial anomaly [14].

Since the separation from $\Sigma$  of the term, proportional to $\Delta g$,
results in redefinition of $\Sigma$, sometimes in the analysis of the data
it is separated, sometimes it is not. In what follows in the main part of
the Lecture I will not separate $\Delta g$ contribution from $\Sigma$, only
sometimes   mentioning how large it could be.

Twist-4 corrections to $\Gamma_{p,n}$
were calculated by Balitsky, Braun and
Koleshichenko (BBK) [15]  using the QCD sum rule method.

BBK calculations were critically analyzed in [16], where it was shown
that there are many possible uncertainties in these calculations: 1)
the main contribution to QCD sum rules comes from the last accounted
term in OPE -- the operator of dimension 8; 2) there is a large
background term and a much stronger influence of the continuum
threshold comparing with usual QCD sum rules; 3) in the singlet case, when
determining the induced by external field vacuum condensates, the corresponding
sum rule was saturated by $\eta$-meson, what is wrong. The next order
term -- the contribution of the dimension 10 operator to the BBK sum
rules was estimated by Oganesian [17]. The account of the dimension-10
contribution to the BBK sum rules and estimation of other uncertainties
results in (see [16]):

\be
b_{p-n} = -0.006 \pm 0.012~GeV^2
\ee
\be
b_{p+n} = -0.035 (\pm 100 \%)~GeV^2
\ee
As is seen from (8), in the nonsinglet case the twist-4 correction is
small ($\la 2\%$ at $Q^2 \ga 5 GeV^2$)
even with the account of the
error. In the singlet case the situation is much worse: the estimate (9)
may be considered only as correct by the order of magnitude.

One may expect that at low $Q^2 <3~GeV^2$ the nonperturbative (higher twist)
corrections  to $\Gamma_{p,n}(Q^2)$  are much larger in absolute values,
than given by (8),(9). This statement follows from the requirement, that at
$Q^2=0~\Gamma_{p,n}(Q^2)$ satisfies the Gerasimov-Drell-Hearn (GDH) sum rule
and a smooth  connection of $\Gamma_{p,n}(Q^2)$ at intermediate $Q^2$ and
those at $Q^2=0$ should exist. (In accord with the GDH sum rule
$\Gamma_{p,n}(0)=0$ and $\Gamma^{\prime}_{p,n}(0)=-\kappa^2_{p,n}/8m^2$,
where $\kappa_{p,n}$ are proton and neutron anomalous magnetic moments --
see [16].)  In [16] the model was suggested, which realizes such smooth
connection. As was demonstrated in [16] the model is in a good agreement
with the  recent experimental data. An interesting feature of the model,
supported by the data, is that the sign of nonperturbetive correction
coincides with the sign of twist-4 terms (7),(8) in the case of proton, but
it is opposite for neutron.

I turn now to comparison of the theory with the recent experimental data.
In Table 1 the recent data obtained by SMC [1],  E154(SLAC) [2] and
HERMES [3] groups are presented.

\vspace{3mm}
\centerline{\large \bf Table ~ 1}

\vspace{2mm}
\begin{tabular}{|c|c|c|c|c|}\hline
&     $\Gamma_p$ & $\Gamma_n$ & $\Gamma_p - \Gamma_n$  & $\alpha_s(5
GeV^2)$\\ \hline
SMC & $0.132 \pm 0.017$ & $ -0.048 \pm0.022$ & $0.181
\pm 0.035$ & $0.270^{+ 0.16}_{-0.40}$\\
\hline
combined & $0.142 \pm 0.011$ & $-0.061 \pm 0.016$ & $0.202 \pm 0.022$
& $0.116^{+ 0.16}_{-0.44}$\\ \hline
E 154(SLAC) & $0.112 \pm 0.014$ & $-0.056 \pm 0.008$ & $0.168 \pm
0.012$ & 0.339 $^{+ 0.052}_{-0.063}$\\ \hline
HERMES & -- & $-0.037 \pm 0.015$ & -- & -- \\ \hline
EJ/Bj sum rules & $0.168 \pm 0.005$ & $-0.013 \pm 0.005$ &
$0.181 \pm 0.002$ & 0.276\\ \hline
\end{tabular}

\vspace{3mm}
In the second line of Table 1 the results of the performed by SMC [1]
combined analysis of SMC [1], SLAC-E80/130 [18], EMC [19] and SLAC-E143
[20] data are given. The data presented in the first three lines of Table 1
refer to $Q^2 = 5~ GeV^2$, HERMES data refer to $Q^2=2.5~GeV^2$.  In all
measurements each range of $x$ corresponds to each own mean $Q^2$.
Therefore, in order to obtain $g_1(x, Q^2)$ at fixed $Q^2$ the authors of
ref.'s [1,2]  used the following procedure.  At some reference scale
$Q^2_0$ ($Q^2_0 = 1 GeV^2$ in [1] and $Q^2_0 = 0.34 GeV^2$ in [2]) quark and
gluon distribution were parametrized as functions of $x$.  (The number of
the parameters was 12 in [1] and 8 in [2]). Then NLO evolution equations
were solved and the values of the parameters were determined from the best
fit at all data points. The numerical values presented in Table 1 correspond
to $\overline{MS}$ regularization scheme, statistical, systematical, as well
as theoretical errors arising from uncertainty of $\alpha_s$ in the
evolution equations, are added in quadratures. The HERMES value of
$\Gamma_n$, measured at $Q^2=2.5~GeV^2$ can be recalculated to $Q^2=5~GeV^2$
using the model [16], matching GDH sum rule at
$Q^2=0$ and asymptotic behavior of $\Gamma_n(Q^2)$. The result is:
$\Gamma_n(Q^2=5~GeV^2)=-0.045\pm 0.015$ (HERMES). In the last line of Table
1 the Ellis-Jaffe (EJ) and Bjorken (Bj) sum rules prediction for $\Gamma_p,
\Gamma_n$ and $\Gamma-\Gamma_n$, correspondingly are given.  The EJ sum rule
prediction was calculated according to (3), where $\Delta s = 0$ , i.e.,
$\Sigma = g^8_A = 0.59$ was put and the last--gluonic term in (3) was
omitted.  The twist-4 contribution was accounted in the Bj sum rule and
included into the error in the EJ sum rule.  The $\alpha_s$ value in the EJ
and Bj sum rules calculation was chosen as $\alpha_s(5~ GeV^2) = 0.276$,
corresponding to $\alpha_s(M_z) = 0.117$ and $\Lambda^{(3)}_{\overline{MS}}
= 360 MeV$ (in two loops). As is clear from Table 1, the data, especially
for $\Gamma_n$ contradict the EJ sum rule. In the last column, the values of
$\alpha_s$ determined from the Bj sum rule are given with the account of
twist-4 corrections.

The experimental data on $\Gamma_p$ presented in Table 1 are not in a
good agreement. Particularly, the value of $\Gamma_p$ given by E154
Collaboration seems to be low: it does not agree with the old data
presented by SMC [21] ($\Gamma_p = 0.136 \pm 0.015$) and E143 [20]
($\Gamma_p = 0.127 \pm 0.011$). Even more strong discrepancy is seen in
the values of $\alpha_s$, determined from the Bj sum rules. The value
which follows from the combined analysis is unacceptably low: the
central point corresponds to $\Lambda^{(3)}_{\overline{MS}} = 15 MeV$!
On the other side, the value, determined from the E154 data seems to be
high, the corresponding $\alpha_s(M_z) = 0.126 \pm 0.009$. Therefore, I
come to a conclusion that at the present level of experimental accuracy
$\alpha_s$ cannot be reliably determined from the Bj sum rule in
polarized scattering.

Table 2 shows the values of $\Sigma$ -- the total nucleon spin
projection carried by $u, d$ and $s$-quarks found from $\Gamma_p$ and $\Gamma_p$
presented in Table 1 using eq.(3). (It was put $g_A = 1.260, g^8_A =
0.59$, the term, proportional to $\Delta g$ is included into $\Sigma$.).

\vspace{3mm}
\centerline{\large \bf Table ~2: ~~The values of $\Sigma$}

\vspace{2mm}
\noindent
\begin{tabular}{|c|c|c|c|c|}\hline
& \multicolumn{2}{|c|}{From $\Gamma_p$}
& \multicolumn{2}{|c|}{From $\Gamma_n$}\\ \hline
& At $\alpha_s(5 GeV^2)=$ & At $\alpha_s(5 GeV^2)$ & At $\alpha_s(5
GeV^2)=$& At $\alpha_s(5 GeV^2)$\\ &$ =0.276$ & given in Table 1 & $ =
0.276$
 & given in Table 1\\ \hline
SMC & 0.296 & 0.294 & 0.294 & 0.296\\ \hline
Comb. & 0.390 & 0.290 & 0.175 & 0.255 \\ \hline
E154 & 0.110(0.17; 0.29) & 0.17(0.24; 0.34) & 0.22(0.28; 0.17) & 0.17
(0.24; 0.13)\\ \hline
HERMES & -- & -- & 0.38(0.26) & at $\alpha_s(2.5 GeV^2) =$ \\
&&&& = 0.337\\ \hline

\end{tabular}

\vspace{3mm}
In their fitting procedure [2] E154 Collaboration used the values $g^8_A
= 0.30$ and $g_A = 1.09$. The values of $\Sigma$ obtained from
$\Gamma_p$ and $\Gamma_n$ given by E154 at $g^8_A = 0.30, g_A = 1.26$ and
$g^8_A = 0.30, g_A = 1.09$ are presented in parenthesis. The value $g^8_A =
0.30$ corresponds to a strong violation of SU(3) flavour symmetry and
is unplausible; $g_A = 1.09$ means a bad violation of isospin and is
unacceptable. As seen from Table 1, $\Sigma$ is seriously affected by
these assumptions. The values of $\Sigma$ found from $\Gamma_p$ and
$\Gamma_n$ using SMC and combined analysis data agree with each other
only,if one takes for $\alpha_s(5 GeV^2)$ the values given in \\
Table 1
($\alpha_s = 0.116$ for combined data), what is unacceptable.

The twist-4 corrections were accounted in the calculations of $\Sigma$  in
Table using eq.'s (8),(9). At $Q^2=5~GeV^2$  they result in increasing of
$\Sigma$ by 0.04 if determined from $\Gamma_n$, at $Q^2=2.5~GeV^2$ (HERMES
data) the twist-4 correction increase $\Sigma$ by 0.06. In the last line in
parenthesis is given the value of $\Sigma$, when higher twist corrections
were found basing on the model matching GDH sum rule and asymptotic
behavior of $\Gamma_n$ [16]. The chosen value of $\alpha_s(2.5
~GeV^2)$=0.337 corresponds to the same
$\Lambda^{(3)}_{QCD}$ $(2~loops)=360~MeV$, as $\alpha_s(5~GeV^2)=0.276$.

To conclude, one may say, that
the most probable value of $\Sigma$ is $\Sigma \approx 0.3\pm 0.1$ .
The contribution of gluons may be estimated as $\Delta g
(1 GeV^2) \approx 0.3$ (see [16]).  Then $\Delta g(5 GeV^2)
\approx 0.6$ and the account of gluonic term in eq.(3) results in increasing
of $\Sigma$ by 0.06.  At $\Sigma = 0.3$ we have $\Delta u = 0.83, \Delta d =
-0.43, \Delta s = -0.1$.

\vspace{3mm}

{\bf 2. The QCD sum rules calculation of $\Sigma$.}

\vspace{2mm}
The quantity $\Sigma$, which has the meaning of proton spin projection,
carried by $u,d,s$ quarks is of a special interest.

An attempt to calculate $\Sigma$ using QCD sum rules in external fields
was done in ref.[22]. Let us shortly recall the idea. The polarization
operator

\begin{equation}
\Pi(p) = i \int d^4 x e^{ipx} \langle 0 \vert T \{ \eta(x),
\bar{\eta} (0) \} \vert 0 \rangle
\end{equation}
was considered, where
\begin{equation}
\eta(x) = \varepsilon^{abc} \Biggl (u^a(x) C \gamma_{\mu} u ^b(x)
\Biggr ) \gamma _{\mu} \gamma_5 d^c(x)
\end{equation}
is the current with proton quantum numbers [23],[24] $u^a, d^b$ are quark
fields, $a,b,c$ are colour indeces. It is assumed that the term
\begin{equation}
\Delta L = j^0_{\mu 5} A_{\mu}
\end{equation}
where $A_{\mu}$ is a constant singlet axial field, is added to QCD Lagrangian.
In the weak axial field approximation $\Pi(p)$ has the form

\begin{equation}
\Pi(p) = \Pi^{(0)} (p) + \Pi^{(1)}_{\mu} (p) A_{\mu}.
\end{equation}
$\Pi^{(1)}_{\mu}(p)$ is calculated in QCD by
OPE at $p^2 < 0, \vert
p^2 \vert \gg R^{-2}_c$, where $R_c$ is the confinement radius. On the
other hand,  using dispersion relation, $\Pi^{(1)}_{\mu} (p)$ is
represented by the contribution of the physical states, the lowest of
which is the proton state. The contribution of excited states is
approximated as a continuum and suppressed by the Borel transformation. The
desired answer is obtained by equalling  these two representations. This
procedure can be applied to any Lorenz structure of $\Pi^{(1)}_{\mu} (p)$ ,
but as was argued in [25,26], the best accuracy can be obtained by
considering the chirality conserving structure $2 p_{\mu} \hat{p} \gamma_5$.

An essential ingredient of the method is the appearance of induced by
the external field vacuum expectation values (v.e.v). The most
important of them in the problem at hand is

\begin{equation}
\langle 0 \vert j^0_{\mu 5} \vert 0 \rangle_A \equiv 3 f^2_0 A_{\mu}
\end{equation}
of dimension 3. The constant $f^2_0$ is related to QCD topological
susceptibility. Using (12), we can write
$$
\langle 0 \vert j^0_{\mu 5} \vert 0 \rangle_A = lim_{q \to 0} ~ i \int~
d^4 xe^{iqx} \langle 0 \vert T \{ j^0_{\nu 5} (x), j^0_{\mu 5} (0) \}
\vert 0 \rangle A_{\nu} \equiv
$$
\begin{equation}
\equiv lim_{q \to 0} P_{\mu \nu} (q) A_{\nu}
\end{equation}
The general structure of $P_{\mu \nu} (q)$ is

\begin{equation}
P_{\mu \nu} (q) = -P_L(q^2) \delta_{\mu \nu} + P_T(q^2) (-\delta_{\mu
\nu} q^2 + q_{\mu} q_{\nu})
\end{equation}
Because of anomaly there are no massless states in the spectrum of the
singlet polarization operator $P_{\mu \nu}$ even for massless quarks.
$P_{T,L}(q^2)$ also have no kinematical singularities at $q^2 = 0$ .
Therefore, the nonvanishing value $P_{\mu \nu} (0)$ comes entirely from
$P_L(q^2)$. Multiplying $P_{\mu \nu} (q)$ by $q_{\mu} q_{\nu}$, in the
limit of massless $u, d, s$ quarks we get

$$
q_{\mu} q_{\nu} P_{\mu \nu} (q) = -P_L (q^2) q^2 = N^2_f
(\alpha_s / 4 \pi)^2 i \int~ d^4 xe^{iqx} \times
$$
\begin{equation}
\times \langle 0 \vert T {G^n_{\mu \nu} (x) \tilde{G}^n_{\mu \nu} (x),
G^m_{\lambda \sigma} (0) \tilde{G}^m _{\lambda \sigma} (0)} \vert 0 \rangle,
\end{equation}
where $G^n_{\mu \nu}$ is the gluonic field strength, $\tilde{G}_{\mu
\nu} = (1/2) \varepsilon_{\mu \nu \lambda \sigma} G_{\lambda \sigma}$.(The anomaly
condition was used, $N_f = 3$.).
Going to the limit $q^2 \to 0$, we have

\begin{equation}
f^2_0 = -(1/3) P_L(0) = \frac{4}{3} N^2_f \chi^{\prime} (0),
\end{equation}
where $\chi(q^2)$ is the topological susceptibility

\begin{equation}
\chi(q^2) = i~ \int d^4 x e^{iqx} \langle 0 \vert T {Q_5 (x),
Q_5 (0)} \vert 0 \rangle
\end{equation}
and $Q_5(x)$ is the topological charge density

\begin{equation}
Q_5(x) = (\alpha_s / 8 \pi)~ G^n_{\mu \nu} (x) \tilde{G}^n _{\mu
\nu} (x),
\end{equation}
As is well known [27], $\chi(0) = 0$ if there is
at least one massless quark. The attempt to find $\chi^{\prime}(0)$ itself
by QCD sum rules failed: it was found [22] that OPE does not converge
in the domain of characteristic scales for this problem. However, it
was possible to derive the sum rule, expressing $\Sigma$ in terms of
$f^2_0$ (14) or $\chi^{\prime}(0)$. The OPE up to dimension $d=7$ was
performed in ref.[22]. Among the induced by the external field v.e.v.'s
besides (14), the v.e.v. of the dimension 5 operator

\begin{equation}
g \langle 0 \vert \sum \limits_{q} \bar{q} \gamma_{\alpha} (1/2) \lambda
^n \tilde{G}^n_{\alpha \beta} q \vert 0 \rangle _A \equiv 3 h_0
A_{\beta}, ~~~ q = u, d, s
\end{equation}
was accounted and the constant $h_0$ was estimated using a special sum
rule,\\  $h_0 \approx 3 \times 10^{-4} GeV^4$ . There were also accounted the
gluonic condensate $d = 4$ and the square of quark condensate $d= 6$
(both times the external $A_{\mu}$ field operator, $d=1$). However, the
accuracy of the calculation was not good enough for reliable
calculation of $\Sigma$ in terms of $f^2_0$: the necessary requirement
of the method -- the weak dependence of the result on the Borel
parameter was not well satisfied.

In [28] the accuracy of the calculation was improved by going to
higher order terms in OPE up to dimension 9 operators. Under the
factorization assumption   -- the saturation of the product of
four-quark operators by the contribution of an intermediate vacuum
state -- the dimension 8 v.e.v.'s were accounted (times $A_{\mu}$):

\begin{equation}
-g \langle 0 \vert \bar{q} \sigma_{\alpha \beta} (1/2) \lambda^n
G^n_{\alpha \beta} q \cdot \bar{q} q \vert 0 \rangle = m^2_0 \langle 0
\vert \bar{q} q \vert 0 \rangle^2,
\end{equation}
where $m^2_0 = 0.8 \pm 0.2 ~GeV^2$
was determined in [28].
In the framework of the same factorization hypothesis the induced by
the external field v.e.v. of dimension 9

\begin{equation}
\alpha_s \langle 0 \vert j^{(0)}_{\mu 5} \vert 0 \rangle_A \langle 0
\vert \bar{q} q \vert 0 \rangle^2
\end{equation}
is also accounted. In the calculation the following expression for
the quark Green function in the constant external axial field was used [26]:

$$
\langle 0 \vert T \{ q^a_{\alpha}(x),~ \bar{q}^b_{\beta} (0) \} \vert 0
\rangle_A = i \delta^{ab} \hat{x}_{\alpha \beta} / 2 \pi^2 x^4 +
$$
$$
+ (1/2 \pi^2) \delta^{ab} (A x) (\gamma_5 \hat{x})_{\alpha \beta} / x^4
- (1/12) \delta^{ab} \delta_{\alpha \beta} \langle 0 \vert \bar{q} q
\vert 0 \rangle +
$$

$$
+(1/72)i\delta^{ab} \langle 0 \vert \bar{q} q \vert 0 \rangle (\hat{x}
\hat{A} \gamma_5 - \hat{A} \hat{x} \gamma_5)_{\alpha \beta} +
$$
\begin{equation}
+ (1/12) f^2_0 \delta^{ab} (\hat{A} \gamma_5)_{\alpha \beta} + (1/216)
  \delta^{ab} h_0 \Biggl [(5/2) x^2 \hat{A} \gamma_5 - (A x) \hat{x}
\gamma_5 \Biggr ]_{\alpha \beta}
\end{equation}
The terms of the third power  in $x$-expansion of
quark propagator proportional to $A_{\mu}$ are omitted in (24),
because they do not contribute to the
 tensor structure of $\Pi_{\mu}$ of interest. Quarks are considered to be
 in the constant external gluonic field and quark and gluon QCD equations
 of motion are exploited (the related formulae are given in [29]). There is
 also an another source of v.e.v. $h_0$  to appear besides the $x$-expansion
 of quark propagator given in eq.(24): the quarks in the condensate absorb
 the soft gluonic field emitted by
 other quark. A similar
 situation takes place also in the calculation of the v.e.v. (23)
 contribution. The accounted diagrams with dimension 9 operators have no
 loop integrations.  There are others v.e.v.  of dimensions $d \leq 9$
 particularly containing gluonic fields.  All of them, however, correspond
 to at least one loop integration and are suppressed by the numerical factor
 $(2 \pi)^{-2}$. For this reason they are disregarded.

The sum rule for $\Sigma$ is given by
$$
\Sigma + C_0 M^2 = -1 + \frac{8}{9 \tilde{\lambda}^2_N} e^{m^2/M^2}
\left \{a^2 L^{4/9} + \right.
$$
\begin{equation}
\left. + 6\pi^2 f^2_0 M^4 E_1 \Biggl (\frac{W^2}{M^2} \Biggr ) L^{-4/9} +
14 \pi^2 h_0 M^2 E_0 \Biggl (\frac {W^2}{M^2} \Biggr ) L^{-8/9} -
\frac{1}{4}~ \frac{a^2 m^2_0}{M^2} - \frac{1}{9} \pi \alpha_s f^2_0~
\frac{a^2}{M^2} \right \}
\end{equation}

Here $M^2$ is the Borel parameter, $\tilde{\lambda}_N$ is defined as
$\tilde{\lambda}^2_N = 32 \pi^4 \lambda^2_N = 2.1 ~ GeV^6$,
$\langle 0 \vert \eta \vert p \rangle = \lambda_N v_p,$
where $v_p$ is proton spinor, $W^2$ is the continuum threshold, $W^2 =
2.5 ~GeV^2$,

\begin{equation}
a = -(2 \pi)^2 \langle 0 \vert \bar{q} q \vert 0 \rangle = 0.55 ~ GeV^3
\end{equation}
$$
E_0(x) = 1 - e^{-x} , ~~~ E_1(x) = 1 - (1 + x)e^{-x}
$$
$L = ln (M/\Lambda)/ ln (\mu/\Lambda),~~~
\Lambda = \Lambda_{QCD} = 200 ~ MeV$ and the normalization point  $\mu$
was chosen $\mu = 1 ~ GeV$.

When deriving (25) the sum rule for the
nucleon mass was exploited what results in  appearance of the first
term, --1, in the right hand side (rhs) of (25). This term absorbs the
contributions of the bare loop, gluonic condensate as well as $\alpha_s$
corrections to them and essential part of terms, proportional to $a^2$ and
$m^2_0 a^2$.
It  must be stressed, that with the account of dimension 9 operators the
OPE series in the calculation of $\Sigma$ is going up to the same order as
OPE in the calculation  of nucleon mass, where in the chirality conserving
sum rule the operators up to dimension 8 were accounted (see Appendix, one
additional dimension in the sum rule for $\Sigma$ comes from the dimension
of external axial field $A_{\mu}$). Therefore, both sum rules are on the
same footing and the procedure of using chirality conserving nucleon sum
rule (A.1) in (25) is legitimate. Otherwise, and this was the drawback of
calculations in [25],[26], the approach is not completely selfconsistent.
The values of the parameters, $a, \tilde{\lambda}^2_N, W^2$
taken above were chosen by the best fit of the sum rules for the nucleon
mass (see [30] and Appendix) performed at $\Lambda = 200 ~ MeV$. It can be
shown, using the value of the ratio $2m_s/ (m_u + m_d) = 24.4 \pm 1.5$  [31]
that $a(1~ GeV) = 0.55 ~ GeV^3$ corresponds to $m_s(1 ~ GeV) = 153 ~ MeV$.
$\alpha_s$ corrections are accounted in the leading order (LO) what results
in appearance of anomalous dimensions. Therefore $\Lambda$ has the meaning
of effective $\Lambda$ in LO.
Its numerical value does not contradicts two loops value of $\Lambda$, used
in Sec.1. (Formally, $\Lambda^{(3)}(2~loops)=360~MeV$ would results to
$\Lambda^{(3)}_{eff}(LO)=250~MeV$.)

The unknown constant $C_0$ in the left-hand
side (lhs) of (25) corresponds to the contribution of inelastic transitions
$p \to N^* \to interaction~ with A_{\mu} \to p$ (and in inverse order). It
cannot be determined theoretically and may be found from $M^2$ dependence of
the rhs of (25) (for details see  [30,32]). The necessary condition of the
validity of the sum rule is $\vert \Sigma \vert \gg \vert C_0 M^2 \vert exp
[(-W^2 + m^2)/M^2]$ at characteristic values of $M^2$  [32]. The
contribution of the last term in the rhs of (25) is negligible.
The sum rule (25) as well as the sum
rule for the nucleon mass is reliable in the interval of the Borel parameter
$M^2$  where the last term of OPE is small, less than $10-15\%$ of the
total and the contribution of continuum does not exceed $40-50\%$ . This
fixes the interval $0.85 < M^2 < 1.4 ~ GeV^2$.The $M^2$-dependence of the
rhs of (25) at $f^2_0 = 3 \times 10^{-2}~ GeV^2$ is plotted in Fig.1. The
complicated expression in rhs of (25) is indeed an almost linear function of
$M^2$ in the given interval! This fact strongly supports the
reliability of the approach. The best values of $\Sigma =
\Sigma^{fit}$ and $C_0 = C^{fit}_0$ are found from the $\chi^2$
fitting procedure

\begin{equation}
\chi^2 = \frac{1}{n} \sum \limits ^{n}_{i=1}~ [\Sigma^{fit} +
C^{fit}_0 M^2_i - R(M^2_i)]^2 = min,
\end{equation}
where $R(M^2)$  is the rhs of (25).

The values of $\Sigma$ as a function of $f^2_0$  are
plotted in Fig.2 together with $\sqrt{\chi^2}$.
In the used above approach  the gluonic
contribution cannot be separated and is included in $\Sigma$. As discussed
in Sec. 1 the experimental value of $\Sigma$ can be estimated
 as $\Sigma = 0.3 \pm 0.1$. Then from Fig.2 we have
$f^2_0 = (2.8 \pm 0.7) \times 10^{-2} ~ GeV^2$ and $\chi^{\prime}(0) = (2.3
\pm 0.6) \times 10^{-3}~  GeV^2$ . The error in $f^2_0$ and $\chi^{\prime}$,
 besides the experimentall error, includes the uncertainty in the sum rule
estimated as equal to the contribution of the last term in OPE (two last
terms in Eq.25) and a possible role of NLO $\alpha_s$ corrections.  At
$f^2_0 < 0.02~ GeV^2$  $\chi^2$ is much worse and the fit becomes unstable.
This allows us to claim (with some care, however,) that $\chi^{\prime} (0)
\geq 1.6 \times 10^{-3} GeV^2$ and $\Sigma \geq 0.05$ from the requirement
of selfconsistency of the sum rule. The $\chi^2$ curve also favours an upper
limit for $\Sigma \la 0.6$.  At $f^2_0 = 2.8 \times 10^{-2}~ GeV^2$  the
value of the constant $C_0$ found from the fit is $C_0 = 0.19 ~ GeV^{-2}$.
Therefore, the mentioned above necessary condition of the sum rule validity
is well satisfied.

Let us discuss the role of various terms of OPE in the sum rules (25)
To analyze it we have considered sum rules (25) for 4 different cases, i.e.
when we take into consideration:
a) only  contribution of the operators  up to d=3 (the term --1 and the term,
proportional to $f^2_0$ in (25));
b) contribution of the operators up to d=5 (the term $\sim h_0$ is added);
c) contribution of the operators up to d=7 (three first terms in (25)),
d) our result (25), i.e. all operators up to d=9.
For this analysis the value of $f^2_0 = 0.03
~GeV^2$ was chosen, but the conclusion  appears to be the same
for all more or less reasonable choice of $f^2_0$. Results of the fit of the
sum rules are shown in  Table 3 for all four cases. The fit is done in the
region of Borel masses $0.9 <M^2 < 1.3~ GeV^2$. In the first column the
values of $\Sigma$ are shown , in the second - values of the parameter C,
and in the third - the ratio $\gamma = \vert \sqrt{\chi^2}/\Sigma \vert$,
which is the real parameter, describing reliability of the fit. From  the
table one can see, that reliability of the fit monotonously improves with
increasing of the number of accounted terms of OPE and is quite satisfactory
in the case $d$

\vspace{2mm}

\begin{center}

\centerline{\large \bf Table 3}

\vspace{2mm}
\begin{tabular}{|c|c|c|c|} \hline
case & $\Sigma$ & $C(GeV^{-2})$ & $\gamma$\\ \hline
a) & -0.019 & 0.31 & $10^{-1}$\\ \hline
b) & 0.031 & 0.3 & $5.10^{-2}$\\ \hline
c) & 0.54 & 0.094 & $9.10^{-3}$\\ \hline
d) & 0.36 & 0.21 & $1.3 \cdot 10^{-3}$ \\ \hline
\end{tabular}

\end{center}

\vspace{2mm}

Recently, the first attempt to calculate
$\chi^{\prime}(0)$ on the lattice was performed [33]. The result is
$\chi^{\prime}(0) = (0.4 \pm 0.2) \times 10^{-3}~  GeV^2$, much below our
value.  However, as mentioned by the authors, the calculation has some
drawbacks and the result is preliminary.

In the papers by Narison, Shore and Veneziano (NSV) [34],[35], an attempt to
find the links between $\Sigma$ and $\chi^{\prime}(0)$  was done. NSV found
that $\Sigma$ is proportional to $\sqrt{\chi^{\prime}(0)}$ and calculated
$\chi^{\prime}(0)$ by QCD sum rules. From my point of view, the approach of
ref.'s [33],[34] is not justifiable. Instead of use of firmly based and
self consistent OPE, as was done above, in [34],[35] the matrix element
$<p\mid Q_5\mid p>$ was saturated by contribution of two operators $Q_5$ and
singlet pseudoscalar operator $\Sigma \bar{q}\gamma_5 q$ -- and the result
was obtained by orthogonalization of the corresponding matrix. I have
doubts  that such procedure can be grounded. The calculation of
$\chi^{\prime}(0)$ by QCD sum rules is not correct, because, as was shown in
[22] by considering in the same problem with account of higher order terms
of OPE, than it was done in [34],[35], the OPE breaks down at the scales,
characteristic for this problem.  I do not believe, that the value
$\chi^{\prime}(0)=(0.5\pm 0.2)\cdot 10^{-3}~GeV^2$ found in [34] is
reliable.

\vspace{3mm}
{\bf 3. Calculation of proton axial coupling constant $g^8_A$ and $g_A$.}

\vspace{2mm}

From the same sum rule (25) it is possible to find $g^8_A$ -- the
proton coupling constant  with the octet axial current, which enters the QCD
formula for $\Gamma_{p,n}$. There are two differences in
comparison with (25):

I. Instead of $f^2_0$ it appears the square $f^2_8$
of the pseudoscalar meson coupling constant with the octet axial
current. In the limit of strict SU(3) flavour symmetry it is equal to
$f^2_{\pi}$,
$f_{\pi} = 133 ~ MeV$. However, it
is known, that SU(3) symmetry is violated and the kaon
decay constant, $f_K \approx 1.22 f_{\pi}$ [6]. In the linear in $s$-quark
mass $m_s$ approximation $f_{\eta} = 1.28 f_{\pi}$.
We put for $f^2_8$ the value $f^2_8 = 2.6 \times 10^{-2}~
GeV^2$, intermediate between $f^2_{\pi}$ and $f^2_{\eta}$ .

2. $h_0$ should be substituted by $m^2_1 f^2_{\pi}$. The constant
$m^2_1$ is determined by the sum rules suggested in [36]. A new fit
corresponding to the values of the parameters used above, was
performed and it was found; $m^2_1 = 0.16~  GeV^2$.

The $M^2$ -dependence of $g^8_A + C_8 M^2$ is presented in Fig.1 and
the best fit according to the fitting procedure (27) at $1.0 \leq M^2
\leq 1.3 ~ GeV^2$ gives

\begin{equation}
g^8_A = 0.65 \pm 0.15, ~~~ C_8 = 0.10 ~GeV^{-2} ~~~\sqrt{\chi^2} = 1.2
\times 10^{-3}
\end{equation}
(The error includes the uncertainties in the
sum rule as well as in the value of $f^2_8$). The obtained value of $g^8_A$
within the errors coincides with $g^8_A = 0.59 \pm 0.02$ [12] found
from the data on baryon octet $\beta$-decays under assumption of strict
SU(3) flavour symmetry and contradicts the hypothesis of bad violation of
SU(3) symmetry in baryon axial octet coupling constants [37].

A similar sum rule with the account of dimension 9 operators can be
derived also for $g_A$ -- the nucleon axial $\beta$-decay coupling
constant. It is an extension of the sum rule found in [25] and has the
form

\begin{equation}
g_A + C_A M^2 = 1 + \frac{8}{9 \tilde{\lambda}^2_N} e^{m^2/M^2} \Biggl
[a^2 L^{4/9} + 2 \pi^2 m^2_1 f^2_{\pi} M^2 - \frac{1}{4} a^2
\frac{m^2_0}{M^2} + \frac{5}{3} \pi \alpha_s f^2_{\pi} \frac{a^2}{M^2} \Biggr ]
\end{equation}
The main term in OPE of dimension 3 proportional to $f^2_{\pi}$
occasionally was cancelled. For this reason the higher order
terms of OPE may be more important in the sum rule for $g_A$ than in the previous
ones. The $M^2$ dependence of $g_A - 1 + C_A M^2$ is plotted in Fig.1,
lower curve; the curve is almost the straight line, as it should be.
The best fit gives

\begin{equation}
g_A = 1.37 \pm 0.10, ~~~ C_A = -0.088~ GeV^{-2}, ~~~~ \sqrt{\chi^2} =
1.0 \times 10^{-3}
\end{equation}
in comparison with the world average $g_A
= 1.260 \pm 0.002$ [6]. The inclusion of dimension 9 operator contribution
essentially improves the result: without it $g_A$ would be about 1.5 and
$\chi^2$ would be much worse.

The work was supported in part by
CRDF Grant RP2-132, INTAS Grant 93-0283, RFFR Grant 97-02-16131 and
Swiss Grant 7SUPJ048716.


\bigskip

\begin{center}
{\bf Appendix}

\vspace{2mm}
{\bf The fit of the sum rules for nucleon mass.}

\end{center}

\setcounter{equation}{0}
\def\theequation{A.\arabic{equation}}

\vspace{3mm}

Since in comparison with previous fit [30] of the sum rules for nucleon
mass the value of QCD parameter was changed now, the new fit was performed.
(In the previous calculations it was used $\Lambda = 100~MeV$, now we take
$\Lambda=200~MeV$.). The sum rules for chirality conserving and chirality
violating parts of the polarization operator $\prod\nolimits^{(0)}(p)$ (6)
defined by (3)  are correspondingly

$$M^6E_2\Biggl (\frac{W^2}{M^2}\Biggr ) L^{-4/9} + \frac{4}{3} a^2 L^{4/9} +
\frac{1}{4}bM^2 E_0 \Biggl (\frac{W^2}{M^2}\Biggr ) L^{-4/9} -$$

\be
- \frac{1}{3}a^2 \frac{m^2_0}{M^2} = \tilde{\lambda}^2_N e^{-m^2/M^2}
\ee

\be
2a M^4 E_1 \Biggl (\frac{W^2}{M^2}\Biggr ) +
\frac{272}{81}~\frac{\alpha_s(M^2)}{\pi} ~\frac{a^3}{M^2} - \frac{1}{12} ab =
m\tilde{\lambda}^2_N e^{-m^2/M^2},
\ee
where

$$b = (2\pi)^2 \langle 0 \mid \frac{\alpha_s}{\pi} G^2_{\mu\nu}\mid 0
\rangle = 0.50~ GeV^4,$$

$$E_2(x) = 1 - (1+x+ \frac{x^2}{2})e^{-x}$$
and the other notations are the same as in (25),(26). Parameters $a$ and
$W^2$  were treated as fitting parameters and it was required that in the
fitting interval $0.8 < M^2 < 1.3~GeV^2$ the quantities $\tilde{\lambda}^2_N$
found from both sum rules (A.1) and (A.2) must be close to one another and
close to a constant, independent of $M^2$. The values of
$\tilde{\lambda}^2_N$, determined from (A.1)  and (A.2) as functions of $a$
(at normalization point $\mu=1~GeV$ and continuum threshold $W^2=2.5~GeV^2$)
are plotted on Fig.3. Two sum rules give the same value of
$\tilde{\lambda}^2_N$ at $a=0.55~GeV^3$. The 10\% variation of $W^2$ does
not change this result.  The $M^2$-dependence of $\tilde{\lambda}^2_N$,
determined from (A.1) and (A.2) at these values of fitting parameters is
shown on Fig.4. As is seen, $\tilde{\lambda}^2_N$ found from two sum rules
agree with one another with accuracy $\sim 3\%$ and their deviation from
constant is less than 5\%. The mean value of $\tilde{\lambda}^2_N$ can be
chosen as $\tilde{\lambda}^2_N=2.1~GeV^6$ $(\mu=1~GeV)$.

\newpage \centerline{\bf Figure Captions}

\vspace{5mm}
\begin{tabular}{lp{12cm}}
{\bf Fig.~1.}  & The $M^2$-dependence of $\Sigma + C_0 M^2$ at $f^2_0 =
3 \times 10^{-2}~
GeV^2$ , eq.25, $g^8_A + C_8 M^2$, and $g_A - 1 + C_A M^2$, eq.29.\\
\\
{\bf Fig.~2.} & $\Sigma$  (solid line, left ordinate
axis) and $\sqrt{\chi^2}$, eq.(27), (crossed line, right ordinate axis).
as a functions of $f^2_0$.\\
\\
{\bf Fig.~3.} & The values of $\tilde{\lambda}^2_N$  as functions of $a$
determined from the sum rules (A.1) -- solid line and (A.2) -- crossed
line.\\
\\
{\bf Fig.~4.} & The $M^2$ -- dependence of $\tilde{\lambda}^2_N$ found from
the sum rules (A.1) -- solid line and (A.2) -- crossed line.

\end{tabular}

\newpage
\begin{figure}
\epsfxsize=10cm
\epsfbox{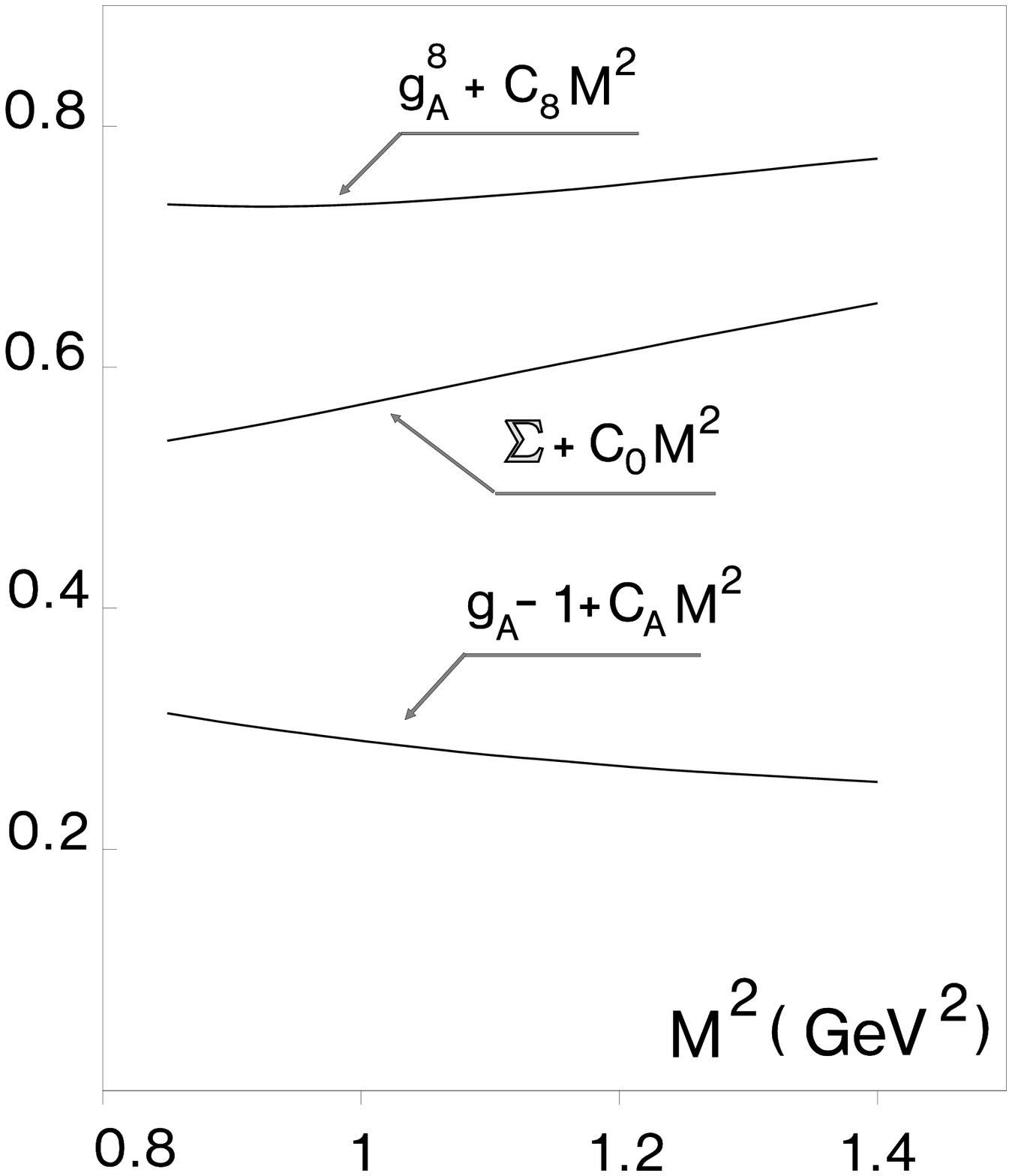}
\caption{}
\end{figure}

\newpage

\begin{figure}
\epsfxsize=10cm
\epsfbox{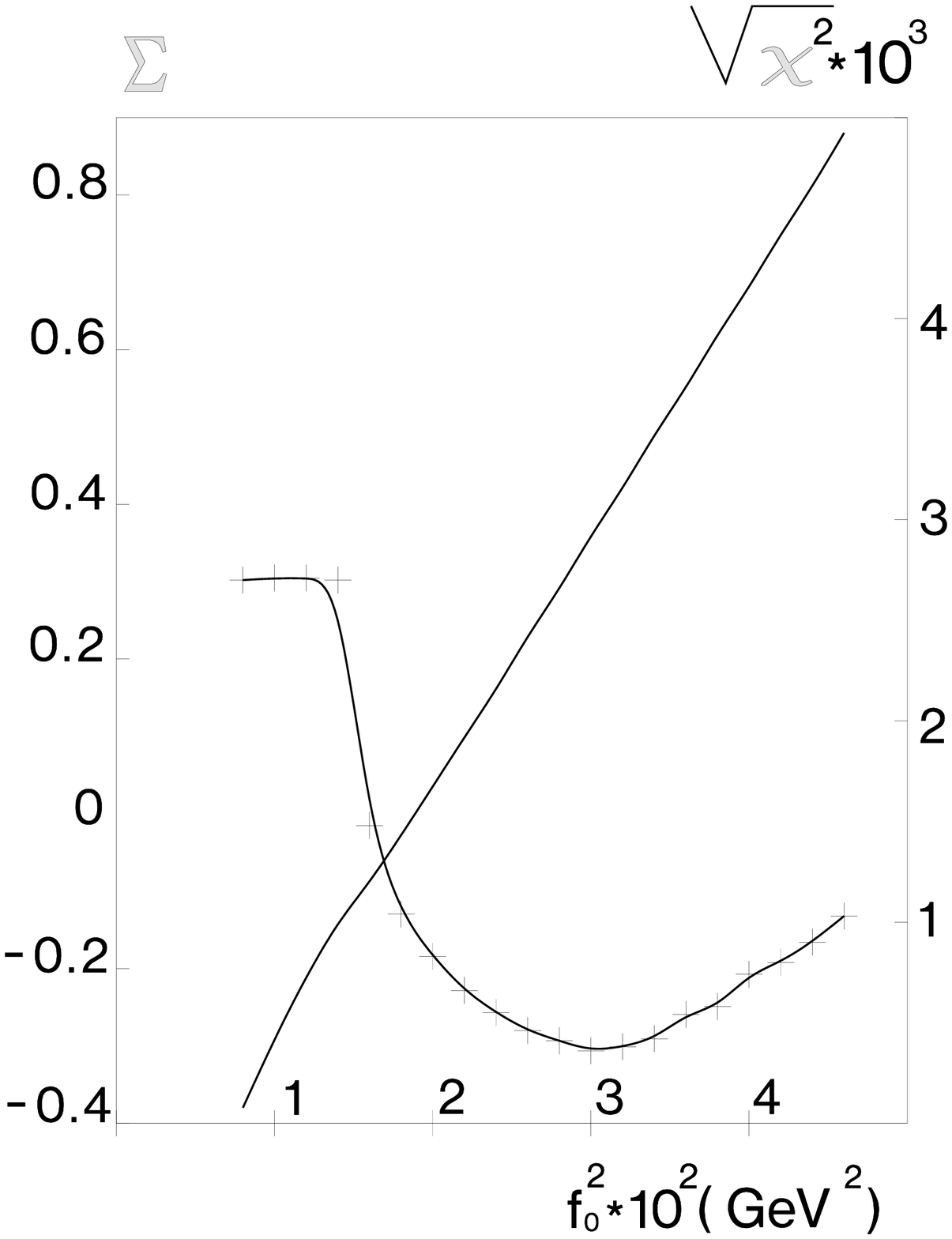}
\caption{}
\end{figure}
\newpage

\begin{figure}
\epsfxsize=10cm
\epsfbox{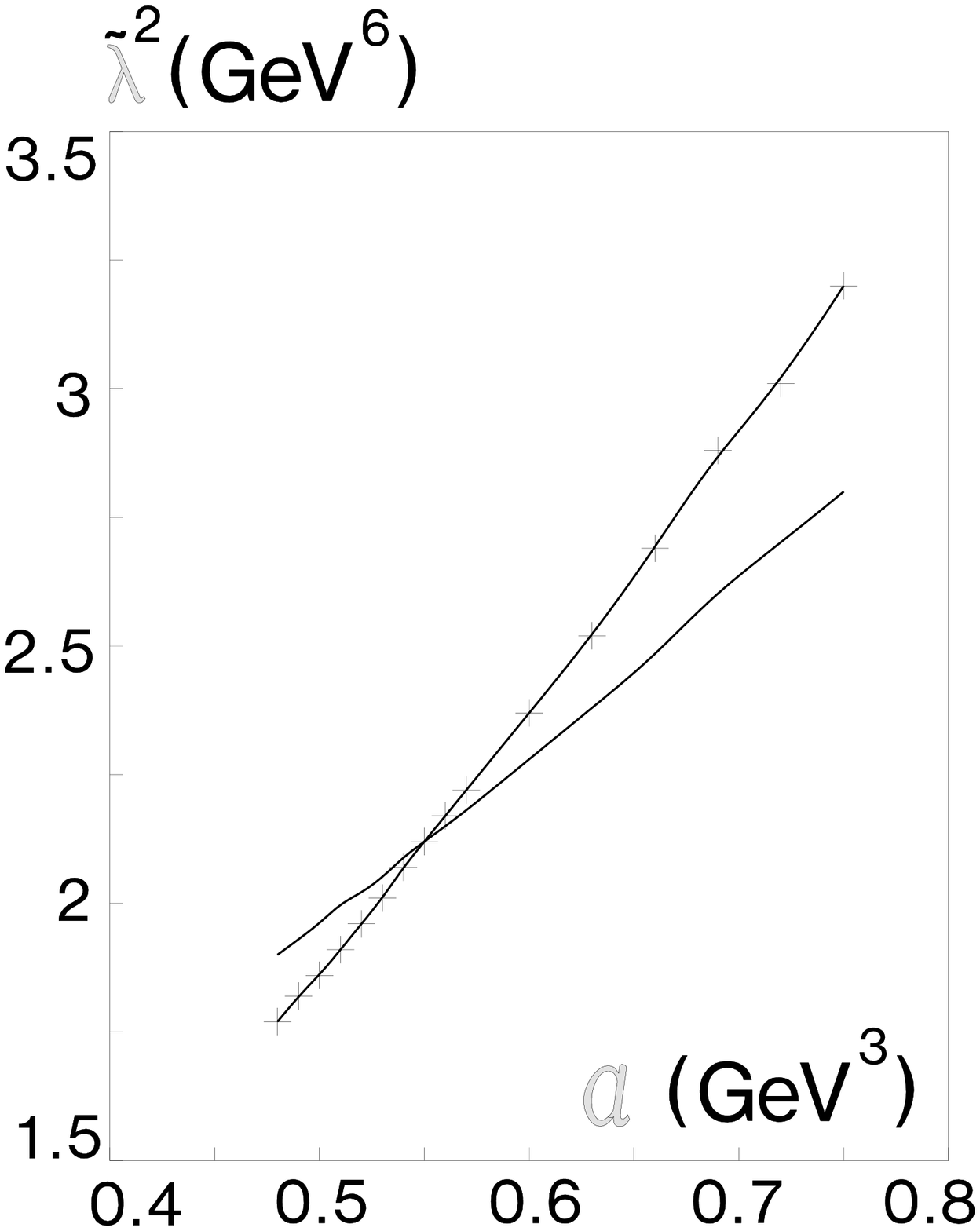}
\caption{}
\end{figure}
\newpage

\begin{figure}
\epsfxsize=10cm
\epsfbox{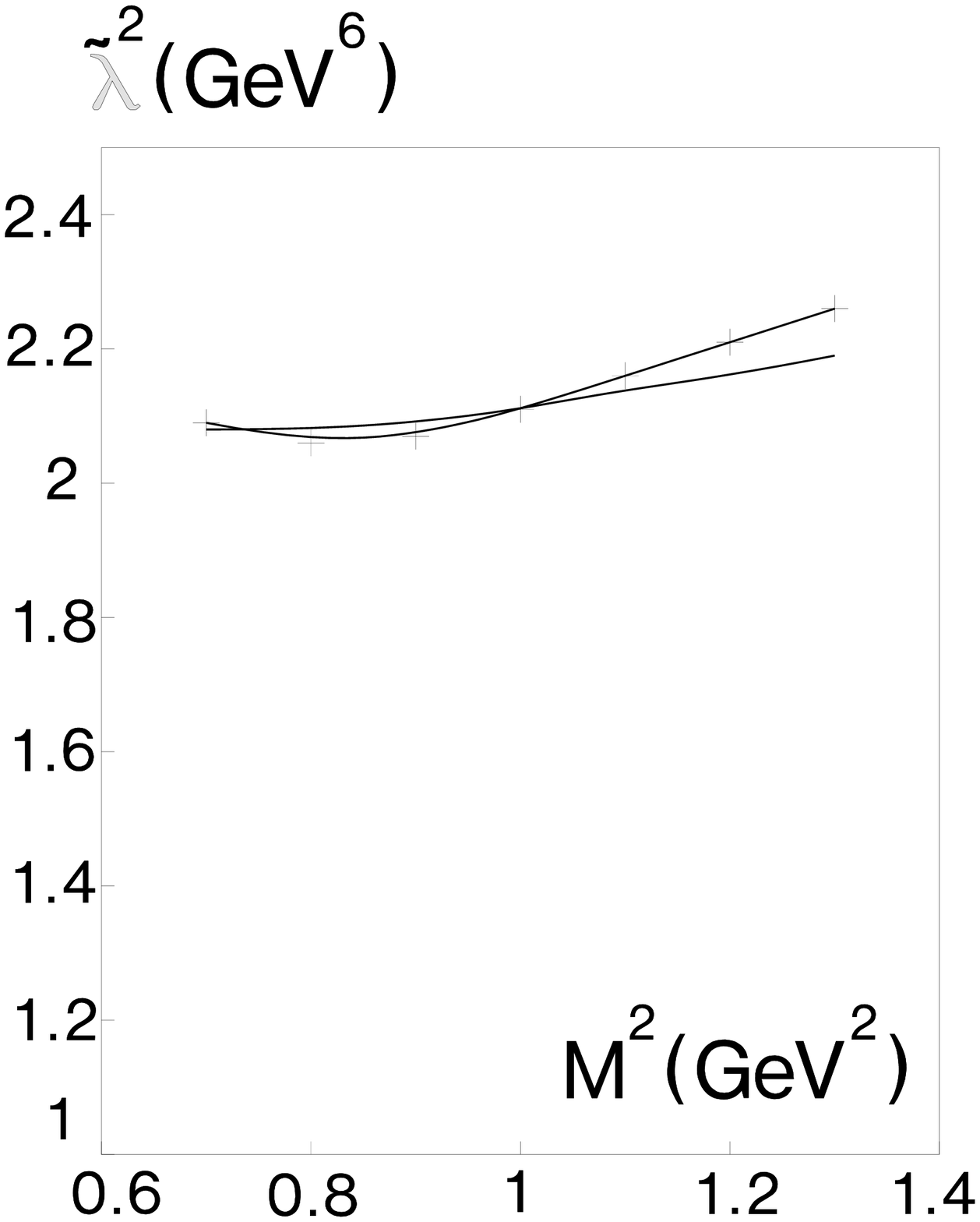}
\caption{}
\end{figure}


\newpage
\begin{thebibliography}{99}
\bibitem{1} D.Adams et al., Phys.Rev. {\bf D56}, 5330 (1997).
\bibitem{2} K.Abe et al. Phys.Lett. {\bf B405}, 180 (1997) .
\bibitem{3} K.Ackerstaff et al., Phys.Lett. {\bf B404}, 383 (1997).
\bibitem{4} M.Anselmino, A.Efremov and E.Leader, Phys.Rep. {\bf 261},
1 (1995).
\bibitem{5} B.L.Ioffe, Int. School of Nucleon Structure, 1st Course:
The Spin Structure of the Nucleon, Erice-Sicily, Aug.1995,
B.Frois and V.Hughes, Eds., New York, Plenum, 1997.
\bibitem{6} R.M.Barnett et al., Particle Data Group, Phys.Rev.
{\bf D54}, 1 (1996).
\bibitem{7} J.Kadaira et al., Phys.Rev. {\bf D20}, 627 (1979);
Nucl.Phys. {\bf B159}, 99 (1979), {\bf 165} 129 (1980).
\bibitem{8} S.A.Larin and J.A.M.Vermaseren, Phys.Lett. {\bf 259}, 345 (1991).
\bibitem{9} S.A.Larin, Phys.Lett. {\bf 334}, 192 (1994).
\bibitem{10} S.A.Larin, T.van Ritbergen and J.A.M.Vermaseren,
Phys.Lett. {\bf B404}, 153 (1997).
\bibitem{11} A.L.Kataev, Phys.Rev. {\bf D50}, 5469 (1994).
\bibitem{12} S.Y.Hsueh et al., Phys.Rev. {\bf D38}, 2056 (1988).
\bibitem{13} R.D.Carlitz, J.C.Collins and A.H.Mueller, Phys.Lett.
{\bf B214}, 229 (1988).
\bibitem{14} S.D.Bass, B.L.Ioffe, N.N.Nikolaev and A.W.Thomas,
J.Moscow Phys.Soc. {\bf 1},
 317 (1991).
\bibitem{15} I.I.Balitsky, V.M.Braun and A.V.Kolesnichenko,
Phys.Lett. {\bf 242},  245 (1990),

Errata {\bf B318},  648 (1993).
\bibitem{16} B.L.Ioffe, Phys.At.Nucl. {\bf 60}, 1866 (1997).
\bibitem{17} A.G.Oganesian, hep-ph/9704435, Phys.At.Nucl., in press.
\bibitem{18} M.J.Alguard et al, Phys.Rev.Lett. {\bf 37},  1261 (1976);
{\bf 41},  70 (1978). G.Baum, ibid {\bf 51},  1135 (1983).
\bibitem{19} J.Ashman et al, Phys.Lett {\bf 206},  364 (1988); Nucl.Phys.
{\bf B328},  1 (1989).
\bibitem{20} K.Abe et al. (SLAC E143 Collaboration),
Phys.Rev.Lett. {\bf 74},  346 (1995).
\bibitem{21} D.Adams et al, Phys.Lett. {\bf B329},  399 (1994).
\bibitem{22} B.L.Ioffe and A.Yu.Khodjamirian, Yad.Fiz. {\bf 55},
3045 (1992).
\bibitem{23} B.L.Ioffe, Nucl.Phys. {\bf B188}, 317 (1981).
\bibitem{24} B.L.Ioffe, Z.Phys. {\bf C 18}, 67 (1983).
\bibitem{25} V.M.Belyaev and Ya.I.Kogan, JETP Lett. {\bf 37}, 730 (1983).
\bibitem{26} V.M.Belyaev, B.L.Ioffe and Ya.I.Kogan, Phys.Lett. {\bf 151B},
290 (1985).
\bibitem{27} R.J.Crewther, Phys.Lett. {\bf 70B}, 349 (1997).
\bibitem{28} B.L.Ioffe and A.G.Oganesian, preprint ITEP-1-98,
hep-ph/9801345, Phys.Rev. D, in press.
\bibitem{29} V.M.Belyaev and B.L.Ioffe, Sov.Phys. JETP {\bf 56}, 493 (1982).
\bibitem{30} B.L.Ioffe and A.V.Smilga, Nucl.Phys. {\bf B232}, 109 (1984).
\bibitem{31} H.Leutwyler, Journ.Moscow Phys.Soc. {\bf 6}, 1 (1996).
\bibitem{32} B.L.Ioffe, Phys.At.Nucl. {\bf 58}, 1408 (1995).
\bibitem{33} G.Boyd et al., hep-lat/9711025.
\bibitem{34}  S.Narison, G.M.Shore and G.Veneziano, Nucl.Phys. {\bf B433},
209 (1995).
\bibitem{35} G.M.Shore and G.Veneziano, preprint CERN-TH/97-206,
hep-ph/9709213.
\bibitem{36} V.A.Novikov et al., Nucl.Phys. {\bf B237}, 525 (1984) .
\bibitem{37} J.Lichtenstadt and H.Lipkin, Phys.Lett. {\bf B353}, 119 (1995).
\end{thebibliography}
\end{document}